\documentclass{ws-mpla}
\usepackage{amsfonts}
\usepackage{amsmath}
\usepackage{amssymb}
\usepackage{hyperref}
\usepackage{graphicx}

\begin{document}

\title{RADIATION AND POTENTIAL BARRIERS OF A 5D BLACK STRING SOLUTION}
\author{MOLIN LIU$^{a}$, HONGYA LIU$^{a}$\thanks{Corresponding author: hyliu@dlut.edu.cn}, LIXIN XU$^{a}$, PAUL S. WESSON$^{b}$}

\address{$^{a}$School of Physics and Optoelectronic Technology, Dalian University of
Technology, Dalian, 116024, P. R. China\\
$^{b}$Department of Physics, University of Waterloo, Waterloo,
Ontario, N2L 3G1, Canada}

\maketitle

%\pub{Received XXX}{Revised XXX}

\begin{abstract}
By using a massless scalar field we examine the effect of an extra
dimension on black hole radiation. Because the equations are
coupled, we find that the structure of the fifth dimension (as for
membrane and induced-matter theory) affects the nature of the
radiation observed in four-dimensional spacetime. In the case of the
Schwarzschild-de Sitter solution embedded in a Randall-Sundrum brane
model, the extension of the black hole along the fifth dimension
looks like a black string. Then it is shown that, on the brane, the
potential barrier surrounding the black hole has a quantized as well
as a continuous spectrum. In principle, Hawking radiation may thus
provide a probe for higher dimensions.
\end{abstract}


\markboth{M. Liu et al.} {Radiation and potential barriers of a 5D
black string solution}

\catchline{}{}{}{}{} %
%%%%%%%%%%%%%%%%%%%%%%%%%%%%%%%%%%%%%%%%%%%%%%%%%%%%%%%%%%%%%%%%%%%%

\ccode{PACS Nos.: 04.70.Dy, 04.50.+h}

\section{Introduction}
Hawking radiation is one of the most important predictions in black
hole physics. A simple proof of it was given by Damour and Ruffini
\cite{ref:Damour} who used a tortoise coordinate to write the radial
part of the Klein-Gordon equation in the form of the
$Schr\ddot{o}dinger-like$ equation, and then found that there is a
potential barrier outside the horizon, and hence implies that black
holes can radiate. There is a long list of research papers on this
in the literature \cite{ref:Page} \cite{ref:Higuchi}
\cite{ref:Grispino}. In the presence of a cosmological constant
$\Lambda$, the Schwarzschild solution is replaced by the
Schwarzschild-de Sitter solution which contains two event horizons
--- an inner black hole horizon and an outer cosmological horizon.
We are living in between these two horizons. It is known that
Hawking radiation also exists in this case \cite{ref:Brady}
\cite{ref:Brevik} \cite{ref:Tian}. However, because $\Lambda$ is
small, its influence on the radiation is expected to be small,
too.

In Kaluza-Klein as well as string/brane theories, it is assumed
that there exists compact or non-compact extra spatial dimensions.
It is known that the 4D Schwarzschild-de Sitter black hole
solution can be embedded into a 5D Ricci-flat manifold with the
following metric \cite{ref:Mashhoon1} \cite{ref:Liu1}
\cite{ref:Wesson1}
\begin{equation} dS^{2}=\frac{\Lambda
\xi^2}{3}\left[f(r)dt^{2}-\frac{1}{f(r)}dr^{2}-r^{2}\left(d\theta^2+\sin^2\theta d\phi^2\right)\right]-d\xi^{2}, \label{eq:5dmetric}%
\end{equation}
where
\begin{equation}
f(r)=1-\frac{2M}{r}-\frac{\Lambda}{3}r^2.\label{f-function}
\end{equation}
The term inside the square bracket is exactly the same
line-element as the 4D Schwarzschild-de Sitter solution.
Therefore, when viewed from a $\xi=constant$ hypersurface, the 4D
line-element represents exactly the Schwarzschild-de Sitter black
hole. However, when viewed from 5D, the horizon does not form a 4D
sphere. It looks like a black string lying along the fifth
dimension. Usually, people call the solution of the 5D equation
$R_{AB}=\Lambda g_{AB}$ the 5D Schwarzschild-de Sitter solution.
Therefore, to distinguish it, we call the solution
(\ref{eq:5dmetric}) a black string, or more precisely, a 5D
Ricci-flat Schwarzschild-de Sitter solution, because
(\ref{eq:5dmetric}) satisfies the 5D vacuum equation $R_{AB}=0$.
So, for (\ref{eq:5dmetric}) there is no cosmological constant when
viewed from 5D. However, when viewed from 4D, there is an
effective cosmological constant. This solution has been studied in
many works \cite{ref:Mashhoon2} \cite{ref:Wesson2} \cite{ref:Liu2}
\cite{ref:Mashhoon3} focusing mainly on the induced cosmological
constant, the extra force and so on. To our knowledge, no one has
studied the radiation of this 5D solution before. Because the
black hole radiation could hopefully be detected in the near
future, and because recent cosmological observations favor the
$\Lambda$ model for dark energy, here we will study the black hole
radiation of this 5D solution.

This paper is organized as follows. In Section 2, we use the
method of separation of variables to analyze the 5D Klein-Gordon
equation. In section 3, we use the R-S type brane model as a
boundary condition to solve the equation for the fifth dimension.
In section 4, we return to the radial equation and obtain the
potential barrier and analyze its spectrum. Section 5 is a
discussion.

%%% ----------------------------------------------------------------------

\section{The Klein-Gordon Equation in the 5D Schwarzschild-de Sitter Solution }

We use a coordinate transformation
\begin{equation}
\xi=\sqrt{\frac{3}{\Lambda}}e^{\sqrt{\frac{\Lambda}{3}}y}\label{replacement}
\end{equation}
and rewrite the 5D metric (\ref{eq:5dmetric}) as
\begin{equation}
dS^{2}=e^{2\sqrt{\frac{\Lambda}{3}}y}\left[f(r)dt^{2}-\frac{1}{f(r)}dr^{2}-r^{2}\left(d\theta^2+\sin^2\theta d\phi^2\right)-dy^{2}\right],\label{eq:5dmetric-y}%
\end{equation}
where {y} is the new coordinate of the fifth dimension. The horizons
of the 5D metric (\ref{eq:5dmetric-y}) are on the hypersurface
$f(r)=0$, which has three solutions
\begin{equation}
f(r)=\frac{\Lambda}{3r}(r-r_{e})(r_{c}-r)(r-r_{o}). \label{re-f function}%
\end{equation}
Here $r_{e}$ is the black hole horizon, $r_{c}$ is the cosmological
horizon, and we are living in between these two horizons. Another
solution $r_{o}=-(r_{e}+r_{c})$ is negative and physically
meaningless.

We consider a massless scalar field $\Phi$ in the 5D spacetime
(\ref{eq:5dmetric-y}). The Klein-Gordon equation for $\Phi$ is
\begin{equation}
\square\Phi=0,\label{Klein-Gorden equation}%
\end{equation}
where $ \square=\frac{1}{\sqrt{g}}\frac{\partial}{\partial
x^{A}}\left(\sqrt{g}g^{AB}\frac{\partial}{\partial{x^{B}}}\right)\label{Dlb}
$  is the 5D d'Alembertian operator. Using the method of separation
of variables, we make the ansatz \cite{ref:Jensen}%
\begin{equation}
\Phi=\frac{1}{\sqrt{4\pi\omega}}\frac{1}{r}R_{\omega}(r,t)L(y)Y_{lm}(\theta,\phi).\label{wave
function}
\end{equation}
Then Eq. (\ref{Klein-Gorden equation}) reduces to three equations:
\begin{equation}\label{5-th-equation}
\frac{d^2L(y)}{dy^2}+\sqrt{3\Lambda}\frac{d L(y)}{dy}+\Omega
 L(y)=0,
\end{equation}
\begin{equation}\label{radius-t-equation}
-\frac{1}{f(r)} r^2\frac{\partial^2}{\partial
t^2}\left(\frac{R_{\omega}}{r}\right)+\frac{\partial}{\partial
r}\left(r^2
 f(r)\frac{\partial}{\partial{r}}\left(\frac{R_{\omega}}{r}\right)\right)-\left[\Omega r^2+l(l+1)\right]\frac{R_{\omega}}{r}=0,
\end{equation}
\begin{equation}\label{sph.equ.}
\frac{1}{\sin\theta}\frac{\partial}{\partial\theta}\left(\sin\theta\frac{\partial{Y_{lm}}}{\partial\theta}\right)+\frac{1}{\sin^2\theta}\frac{\partial^2
Y_{lm}}{\partial{\phi^2}}=-l(l+1)Y_{lm},
\end{equation}
where $\Omega$ and $\l$ are the two separation constants to be
determined later, after the application of a boundary condition.

\section{The Wave Function $L(y)$}
The wave function $L(y)$ is governed by Eq. (\ref{5-th-equation})
which is a linear second-order differential equation with constant
coefficients and can be rewritten as
\begin{equation}
 \frac{d^2}{dy^2}[e^{\frac{\sqrt{3\Lambda}}{2}y}L(y)]+(\Omega-\frac{3\Lambda}{4})
 [e^{\frac{\sqrt{3\Lambda}}{2}y}L(y)]=0.\label{5th-equation}\\
 \end{equation}
Then, for the three cases $\Omega>\frac{3\Lambda}{4}$,
$\Omega=\frac{3\Lambda}{4}$, and $\Omega<\frac{3\Lambda}{4}$,  we
obtain
\begin{equation}
L(y)=\left\{
\begin{array}{c}
C e^{-\frac{\sqrt{3\Lambda}}{2}y}\cos\left[\sqrt{\Omega-\frac{3}{4}\Lambda}(y-y_{0})\right],{\ \ \ \ \ }{\ \ \ }\text{\ for }\Omega >\frac{3\Lambda}{4},\\
\left(C_{1}+C_{2}y\right)e^{-\frac{\sqrt{3\Lambda}}{2}y},{ \ \ \ \ \ \ \ \ \ \ \ \ \ \ }{\ \ \ \ \ \ \ \ \  \ \ \ \ }\text{\ for }\Omega=\frac{3\Lambda}{4},\\
C_{3}e^{\frac{-\sqrt{3\Lambda}+\sqrt{3\Lambda-4\Omega}}{2}y}+C_{4}e^{\frac{-\sqrt{3\Lambda}-\sqrt{3\Lambda-4\Omega}}{2}y},\text{\ \ \ \ for }\Omega<\frac{3\Lambda}{4},\\

\end{array}
\right.\label{5-d-function}
\end{equation}
where $(C, y_{0})$, $(C_{1}, C_{2})$ and $(C_{3}, C_{4})$ are the
three pairs of integration constants for the three equations in
Eqs. (\ref{5-d-function}), respectively.

Because $L(y)$ represents the wave function along the
$y$-direction for a massless scalar particle and $|L(y)|^{2}$
represents the probability of finding the particle, a boundary
condition should be used to make $L(y)$ finite everywhere. A
simple way is to use the R-S two-brane model \cite{ref:Randall1}
\cite{ref:Randall2}, in which the fifth dimension is a line
segment. The length of this line could be very small
\cite{ref:Randall1}, or could be very large \cite{ref:Randall2}.
We suppose that the two branes are at $y=0$ and $y=y_{1}$,
respectively, and that we live on the $y=0$ brane. Then $L(y)$
should reach its maximum at $y=0$. It is shown that the first
equation in Eqs. (\ref{5-d-function}) gives a discrete spectrum
for $L(y)$, and the second and third equations give continuous
spectra. We discuss them in the following subsections.

\subsection{Discrete Spectra $(\Omega
>\frac{3\Lambda}{4})$}
\begin{figure}[tbh]
\centering
\includegraphics[width=3.5in]{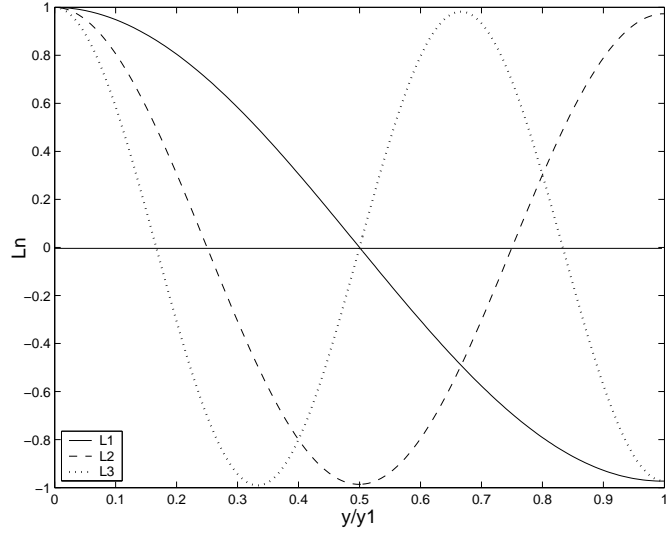}
\caption{The first three eigenfunctions of $L_{n}(y)$:
$L_{1}$(solid), $L_{2}$(dashed), and $L_{3} $(dotted) with $C=1$ and
$\Lambda=10^{-3}$. For illustration, a very large 5th dimension with
$y_{1}=1$ is used here.}\label{fig:quan-state1}
\end{figure}

For the first solution of Eqs. (\ref{5-d-function}), it is
expected that the wave function $L(y)$ should reach its maximum on
our brane $y=0$, and hence $y_{0}=0$ is obtained. Then we expect
that the cosine function in Eqs. (\ref{5-d-function}) looks like a
standing wave between the two branes, so we have
\begin{equation}
y_{1}\sqrt{\Omega-\frac{3}{4}\Lambda}=n\pi,   {\ \ \ } n=1,2,3\cdots,\\
\label{y1}
\end{equation}
and the quantized discrete spectra for $\Omega$ and $L(y)$ are
obtained as follows:
\begin{eqnarray}
  \Omega_{n} &=& \frac{n^2\pi^2}{y_{1}^2}+\frac{3}{4}\Lambda, {\ \ \ } n=1,2,3\cdots,\label{quan-omega} \\
  L_{n}(\phi) &=& C e^{-\frac{\sqrt{3\Lambda}}{2}y}\cos\left(n\pi\frac{y}{y_{1}}\right),{\ \ \ } n=1,2,3\cdots,\label{quan-L}
\end{eqnarray}
It is easy to verify that the corresponding eigenvalue equation and
the operator for $\Omega_{n}$ are
 \begin{equation}
 \hat{\Omega}L_{n}(y)=\Omega_{n}L_{n}(y),{\ \ \ }\hat{\Omega}=-\frac{d}{dy}\left(\frac{d}{dy}+\sqrt{3\Lambda}\right).\label{5-d-energy-vector}
 \end{equation}
For $n=1$, the first (ground) eigenfunction is
$L_{1}(y)=Ce^{-\frac{\sqrt{3\Lambda}}{2}y}\cos(\frac{\pi{y}}{y_{1}})$.
For $n=2$ and $n=3$, the second and third eigenfunctions are
$L_{2}=Ce^{-\frac{\sqrt{3\Lambda}}{2}y}\cos(\frac{2\pi{y}}{y_{1}})$
and
$L_{3}(y)=Ce^{-\frac{\sqrt{3\Lambda}}{2}y}\cos(\frac{3\pi{y}}{y_{1}})$,
respectively. These three eigenfunctions are plotted in Fig.
\ref{fig:quan-state1}. It is clear that $L_{n}(y)$ starts from the
value $L_{n}(y)=1$ but does not end exactly at the value
$L_{n}(y)=\pm1$. This is due to the effect of the exponential factor
$e^{-\frac{\sqrt{3\Lambda}}{2}y}$
 in $L_{n}(y)$ in Eq. (\ref{quan-L}).

\subsection{Continuous Spectra $(\Omega
\leq\frac{3\Lambda}{4})$}

\begin{figure}[tbh]
\centering
\includegraphics[width=3.5in]{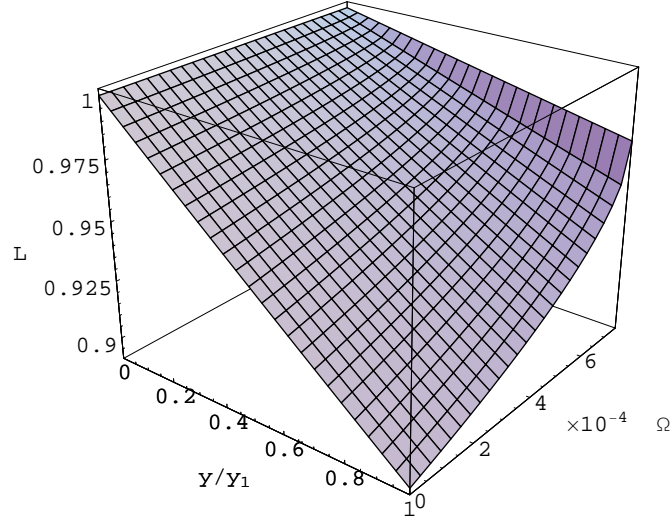}
\caption{The continuous wave function L(y,$\Omega$) for
$\Omega<\frac{3}{4}\Lambda$ with $C_{3}=-1$, $C_{4}=2$,
$\Lambda=10^{-3}$ and $y_{1}=1$ (a very large 5th
dimension).}\label{fig:less-than0.75}
\end{figure}

For $\Omega\leq\frac{3}{4}\Lambda$, the wave function $L(y)$ is
given by the last two equations in Eqs. (\ref{5-d-function}). We
rewrite these as follow:
\begin{equation}
L(y)=\left\{
\begin{array}{c}
\left(C_{1}+C_{2}y\right)e^{-\frac{\sqrt{3\Lambda}}{2}y},\text{ \ \ }\text{ \ \ }\text{ \ \ \ }\text{ \ \ \ \ \ \ }\text{ \ \ \ \ \ }\text{ \ for }{\ \ }\Omega=\frac{3}{4}\Lambda,\\
C_{3}e^{\frac{-\sqrt{3\Lambda}+\sqrt{3\Lambda-4\Omega}}{2}y}+C_{4}e^{\frac{-\sqrt{3\Lambda}-\sqrt{3\Lambda-4\Omega}}{2}y},\text{ \ for }{\ \ }\Omega<\frac{3}{4}\Lambda.\\
\end{array}
\right.\label{y1-function}
\end{equation}
Because $\Omega$ is continuous, the wave function $L(y)$ is also
continuous. Here, just as the discrete case, we require that the
wave function $L(y)$ starts from the maximum on the $y=0$ brane
and decreases as $y$ increases. Under this requirement the case
$\Omega=\frac{3}{4}\Lambda$ contains two modes: for ($C_{1}=1,
C_{2}=0$), $L(y)$ decreases slightly because of the small
$\Lambda$. For ($C_{1}=0, C_{2}<0$), it decreases almost linearly
to $y$. The $\Omega<\frac{3}{4}\Lambda$ case is a little more
complicated because both $y$ and $\Omega$ can vary. The
significant variation can be illustrated in Fig.
\ref{fig:less-than0.75} where $C_{1}=-1$, $C_{2}=2$ and $y_{1}=1$
are chosen.

\section{The Radial Equation and Quantized Potential}
Now we return to the radial Eq. (\ref{radius-t-equation}). Here $t$
can be eliminated by using
 \begin{equation}
R_{\omega}(r,t)\rightarrow\Psi_{\omega l}(r) e^{-i\omega t},
 \end{equation}
 and hence Eq. (\ref{radius-t-equation}) is rewritten as
 \begin{equation}
 \left[-f(r)\frac{d}{dr}(f(r)\frac{d}{dr})+V(r)\right]\Psi_{\omega
 l}(r)=\omega^2\Psi_{\omega l}(r),\label{radius equ. about r}
 \end{equation}
where the potential is given by%
\begin{equation}
V(r)=f(r)\left[\frac{1}{r}\frac{df(r)}{dr}+\frac{l(l+1)}{r^2}+\Omega\right].\label{potential
of r}
\end{equation}

Furthermore, we introduce the tortoise coordinate%
\begin{equation}
x=\frac{1}{2M}\int\frac{dr}{f(r)}.\label{tortoise }
\end{equation}
Integrating this equation shows that $x$ can be expressed explicitly in the following form:%
\begin{equation}
x=\frac{1}{2M}\left[\frac{1}{2K_{e}}\ln\left(\frac{r}{r_{e}}-1\right)-\frac{1}{2K_{c}}\ln\left(1-\frac{r}{r_{c}}\right)+\frac{1}{2k_{o}}\ln\left(1-\frac{r}{r_{o}}\right)\right],\label{tortoise
of gravitaion surface}
\end{equation}
where%
\begin{equation}
K_{i}=\frac{1}{2}\left|\frac{df}{dr}\right|_{r=r_i},
\end{equation}
explicitly as
\begin{equation}
    K_{e}=\frac{(r_{c}-r_{e})(r_{e}-r_{o})}{6r_{e}}\Lambda,
\end{equation}
\begin{equation}
    K_{c}=\frac{(r_{c}-r_{e})(r_{c}-r_{o})}{6r_{c}}\Lambda,
\end{equation}
\begin{equation}
   K_{o}=\frac{(r_{o}-r_{e})(r_{c}-r_{o})}{6r_{o}}\Lambda.
\end{equation}
Under the tortoise coordinate transformation (\ref{tortoise }), the
radial Eq. (\ref{radius equ. about r}) reads%
\begin{equation}
\left[-\frac{d^2}{dx^2}+4M^2V(r)\right]\Psi_{\omega
l}(x)=4M^2\omega^2\Psi_{\omega l}(x).\label{radius-equation}
\end{equation}
This is a $Schr\ddot{o}dinger-like$ equation which describes a
one-dimensional wave propagating between the black hole horizon
$r_{e}$ and the cosmological horizon $r_{c}$. The potential barrier
$V(r)$ determines the reflection and the transmission coefficients
of the scalar massless particles, and the non-zero reflection
coefficient indicates the existence of Hawking radiation.

\begin{figure}[tbh]
\centering
\includegraphics[width=3.5in]{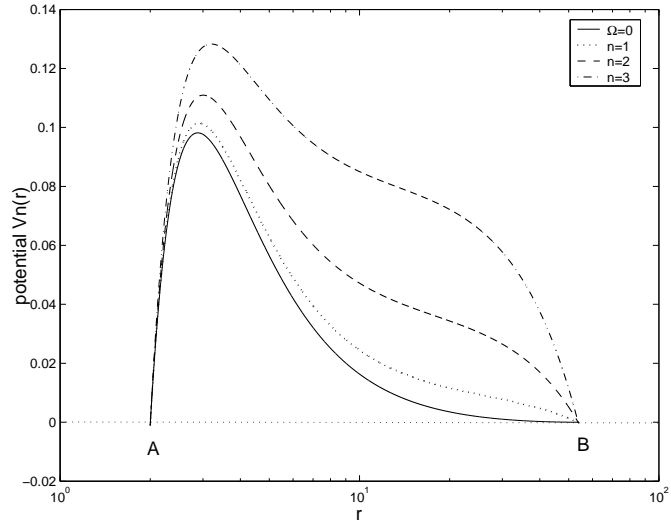}
\caption{The first three potential barriers $V_{n}(r)$ with $n=1$
(dotted), $n=2$ (dashed), and $n=3$ (dash-dot) with $l=1$, $M=1$,
$\Lambda=10^{-3}$ and $y_{1}=10^{3/2}$ (a very large 5th dimension).
The potential barriers of the corresponding 4D solution ($\Omega=0$)
is also plotted for comparison. The black hole horizon (point A) has
$r_{e}\sim2$ and the cosmological horizon (point B) has
$r_{c}\sim54$ and the potential tends to zero quickly as
$x\rightarrow\pm\infty.$}\label{fig:quan-potential-r}
\end{figure}

Now we consider the potential barrier $V(r)$ (\ref{potential of
r}) where $\Omega$ can take a continues value if
$\Omega\leq\frac{3}{4}\Lambda$, or a discrete value if
$\Omega>\frac{3}{4}\Lambda$. In the latter discrete case, $V(r)$
takes the form
\begin{equation}\label{quan-potential}
    V_{n}(r)=f(r)\left[\frac{1}{r}\frac{df(r)}{dr}+\frac{l(l+1)}{r^2}+\frac{n^2\pi^2}{y_{1}^2}+\frac{3}{4}\Lambda\right].
\end{equation}
From this equation it is obvious that if $y_{1}$ is very small,
the potential barrier $V_{n}$ will be very large. However, the
potential barrier for the 4D Schwarzschild-de Sitter solution
corresponds to $\Omega=0$. To make the 5D and 4D cases comparable,
a very large 5th dimension for $y_{1}$ have to be chosen. This
enables us to plot the first three quantized potential barriers
$V_{n}(r)$, as well as the 4D ($\Omega=0$) case, in Fig.
\ref{fig:quan-potential-r}.

\section{Discussion}
The 4D Schwarzschild-de Sitter black hole solution can be embedded
into the 5D Ricci-flat solution (\ref{eq:5dmetric}). We have
analyzed the nature of a massless scalar particle moving in the 5D
solution (\ref{eq:5dmetric}), by solving the corresponding 5D
Klein-Gordon Eq. (\ref{Klein-Gorden equation}) under the
assumption of separability. We have assigned boundary condition in
the 5th dimension by using a two-brane Randall-Sundrum model. The
separation constant $\Omega$ in Eq. (\ref{5-th-equation}) for the
extra dimension plays a critical role, since its value leads to
both quantized and continuous spectra. When $\Omega$ is inserted
into the radial Eq. (\ref{radius-t-equation}), we find that by
using the tortoise coordinate (\ref{tortoise }) we obtain a
one-dimensional $Schr\ddot{o}dinger-like$ Eq.
(\ref{radius-equation}) in which quantized potential barriers
differ significantly from the usual 4D case. (The latter is
recovered for $\Omega=0$, whereas in the 5D case $\Omega$ takes
discrete values for $\Omega>3\Lambda/4$ and continuous values for
$\Omega\leq3\Lambda/4$.) Now in a usual language, as the potential
barrier becomes higher, the reflection of waves become more
efficient and the black hole radiation is more intensive. Thus we
conclude that the existence and structure of the 5th dimension can
in principle be investigated by data about the 4D Hawking
radiation around black hole.

Here we should emphasize that in the brane-world scenario, all
standard particles and forces, except gravity, are confined on the
branes. This implies that the 5D energy and pressure densities
(including a possible positive or negative cosmological constant)
are singular at the branes. Usually, this is treated by using
their 4D counterparts (which are finite) multiplied by a delta
function. It is known that a delta function can be approximately
described by a series of periodic function. In this case, although
the delta function itself is singular at the brane, each of its
component function is not necessarily be singular there. In other
words, if we only focus on a single mode of the series, the wave
function for this mode may reach the maximum value but keep smooth
and finite at the brane. With this purpose in mind, we have used
the RS2-type brane model to give the required boundary conditions.
This is because in the RS2 brane model, the 5th dimension $y$
could be very large and the second brane could be pushed far away
even to infinity. This is just what we want in our paper because
the potential barrier $V_{n}(r)$ tends to its 4D GR value for
$y_{1}\rightarrow\infty$ as can be seen from Eq.
(\ref{quan-potential}). Meanwhile, the 4D spacetime on the brane
in our case takes exactly the same form as the 4D Schwarzschild-de
Sitter solution which is 5D empty. Therefore, there would be no
brane if we do not consider the contribution of the wave function
of the scalar field. The wave function, especially $L(y)$, plays
the role to provide a kind of brane which is similar to the one in
the RS2 model. For instance, a suitable superposition of some of
the quantized or/and continuous components of $L(y)$ may provide a
wave function which is very large at $y=0$ and drops rapidly for
$y\neq0$ and thus forms a practical brane at the $y=0$
hypersurface.

\section{Acknowledgments}

We are grateful to Feng Luo for valuable discussions. This work
was supported by NSF(10573003) and NBRP(2003CB716300) of P. R.
China and by NSERC of Canada. Xu was supported in part by DUT
893321.

\end{document}